\documentclass[journal = ancac3, layout = traditional]{achemso} 

\setkeys{acs}{usetitle=true}
\setkeys{acs}{maxauthors=20}

\usepackage{color}
\usepackage{xcolor}
\usepackage{graphicx}
\usepackage{mathrsfs}
\usepackage{bm}
\usepackage{soul}
\usepackage{amssymb}
\usepackage{amsmath}
\usepackage{physics}
\usepackage{array}
\usepackage{upgreek}

\title[Tandem Solar Cell]{Two-Terminal Tandem Solar Cells based on Perovskite and Transition Metal Dichalcogenides}

\author{Harishankar Suman}
\affiliation{School of Basic Sciences, Indian Institute of Technology Bhubaneswar, Jatni, 752050 Khurda, India}

\author{Avijit Kumar}
\email{avijitkumar@iitbbs.ac.in}
\affiliation{School of Basic Sciences, Indian Institute of Technology Bhubaneswar, Jatni, 752050 Khurda, India}


\begin{document}

\begin{abstract}
Perovskite solar cells have shown power conversion efficiencies (PCE) comparable to cystalline silicon solar cell despite involving low-temperature, solution based synthesis processes outside clean room environment. As the theoretical PCE of a perovskite solar cell with band gap 1.55 eV is capped to 33 \% due to Shockley–Queisser limit, tandem configurations are being investigated to go beyond this limit. Here, we propose a two-terminal (2T) tandem solar cell structure consisting of perovskite and multilayer transition metal dichalcogenide as the absorber layers of the top and the bottom subcells, respectively and investigate their performance parameters using Solar Cell Capacitance Simulator-1 Dimension (SCAPS-1D) software package. We demonstrate that the 2T tandem solar cell consisting of CH$_3$NH$_3$PbI$_3$ with band gap 1.55 eV and MoTe$_2$ with bandgap 1.1 eV shows PCE of maximum 35.3 \% under AM 1.5 G illumination. This work motivates experimental realization of such solar cells for further investigation. 

\end{abstract}

\maketitle

\section*{Introduction}

Perovskite based solar cells have seen significant rise in solar efficiency in less than a decade since 2012 \cite{Kojima2009, Park2020, nrel}. They are attractive as they involve  low-temperature, solution and vapor-phase based fabrication processes using low cost precursors \cite{Yang2017, Momblona2016} without the requirement of cleanroom facilities compared to fabrication of conventional silicon based solar cells. The high power conversion efficiency (PCE) in the perovskite solar cells are attributed to higher charge carriers mobility \cite{Herz2017}, long carrier diffusion lengths \cite{Xing2013}, high absorption coefficients \cite{DeWolf2014} and tuneable energy band gap \cite{Ju2018}. These properties have led to a remarkable PCE of greater than 25 \% in case of single junction perovskite solar cells \cite{Yoo2021, nrel}. Even though perovskite based solar cells are promising, their efficiency is capped by the theoretical Shockley-Queisser limit (PCE = 33 \% for perovskite of band gap 1.34 eV \cite{Shockley1961}). Therefore, tandem solar cell architectures are being investigated to surpass this limit \cite{Leijtens2018,Zhang2020,Li2020}.

Tandem solar cell structure consists of two junctions in series with higher band gap material forming the top junction and lower band gap material forming the bottom junction \cite{Leijtens2018, Zhang2020}. Here, we are interested in two-terminal (2T) tandem solar cell which require less number of functional layers compared to the four-terminal tandem solar cells. This leads to low optical and electrical losses resulting into economical solar cells. Tunable band gap of the perovskites allows to synthesize all-perovskite tandem solar cells with maximum experimental reported PCE of 24.8 \% \cite{Lin2019}. However, all-pervoskite tandem solar cells suffer from stability problems of low-band gap, bottom layer perovskite along with the migration and mixing of extremely mobile ions within the top and the bottom layers \cite{Zhang2020}. Similarly, tandem solar cell based on perovskites and copper indium gallium selenide (CIGS) also suffer from major challenge of commercialization due to scarcity of gallium and indium \cite{Fthenakis2012}. 

Another notable example of 2T tandem solar cell consists of perovskite as top layer and silicon substrates as the bottom layer. Recently, Al-Ashouri et al. \cite{AlAshouri2020} reported a maximum efficiency of $\sim$30 \% for such a tandem solar consisting of Cs$_{0.05}$(FA$_{0.77}$MA$_{0.23}$)$_{0.95}$Pb(I$_{0.77}$Br$_{0.23}$)$_3$ as the top subcell and silicon heterojunction as the bottom subcell. Such high efficiency has been attributed to a fast and efficient hole-extracting self-assembled monolayer which minimizes non-radiative carrier recombination and also forms lossless interface between the indium tin oxide (ITO) and the perovskite layer. However, the solid and rigid properties of silicon substrates will render perovskite-Si based  tandem structure unsuitable for  flexible substrate solar cells applications for which perovskites based solar cells have gained some attraction.  



Here, we propose a 2T tandem solar cell structures consisting of perovskite and Transition Metal Dichalcogenides (TMDCs) as absorber layers for the top and bottom subcells, respectively and explore its performance parameters. TMDCs are layered crystals which exhibit diverse electrical properties ranging from metallic, semiconducting, ferromagnetic to superconducting phases \cite{Roldan2014, Manzeli2017}. TMDCs such as MoTe$_2$, MoS$_2$, MoSe$_2$, WS$_2$, and WSe$_2$ are semiconductors and possess excellent optoelectronic properties including large absorption coefficient over broad wavelength spectrum \cite{JaegerWaldau1994, Tsai2013}. Experimental demonstration of multilayer MoS$_2$ \cite{Wi2014} and MoSe$_2$ \cite{Memaran2015} as photovoltaics exhibits potential for their use in solar cell applications. Further, they are highly stable in ambient condition and can be grown using various techniques such as chemical vapor transport \cite{Reale2016}, chemical vapour deposition \cite{Mandyam2020}, and hydrothermal technique \cite{Chen2001}. However, their applications in solar cells require synthesis of multilayer, large area  TMDCs to be improved and optimized for large area applications.  

The band-gap tuneability, excellent optical properties of both perovskite and TMDCs makes it imperative to investigate performance parameters of 2T tandem solar cell consisting of perovskite and TMDCs as absorber layers of the top and bottom subcells. We use Solar Cell Capacitance Simulator-1 Dimension (SCAPS-1D; version 3.3.08) software package for simulation and study of the single junction and 2T tandem structure solar cells \cite{Burgelman2000}. We find that the maximum PCE of the optimized tandem structure is 35.3 \% for perovskite of band gap 1.55 eV and MoTe$_2$ having band gap of 1.1 eV. Flexibility of the TMDCs and high absorption coefficient will make this structure attractive for various large-scale applications where thin solar cells can be manufactured in flexible substrates through roll-to-roll-based technologies. Our results provide a motivation for experimental realization of such solar cells for further investigation.

\section*{Results and Discussion}


\begin{figure}[h!]
  \includegraphics[width=0.70\textwidth]{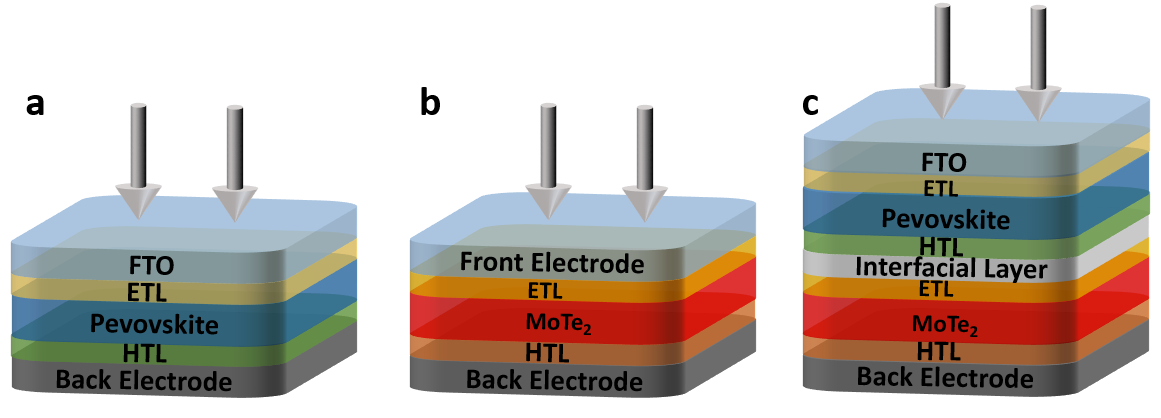}
  \caption{\textbf{a} shows single junction perovskite solar cell structure with various layers. \textbf{b} shows single junction MoTe$_2$ solar cell structure with various layers. \textbf{c} shows 2T tandem solar cell structure containing perovskite and MoTe$_2$ as the absorber layers of the top and bottom subcells, respectively. The arrows indicate that the light is illuminating them from the top.}
  \label{fig:fig1}
\end{figure}

Figure \ref{fig:fig1} shows schematics of two single junction solar cells containing perovskite and MoTe$_2$ as absorber layers and a 2T tandem solar cell structure containing both absorber layers. The light is incident through the top transparent layer as indicated by the arrows and the bottom layer corresponds to the back electrode. Fig. \ref{fig:fig1}(a) shows arrangement of different layers in single junction perovskite solar cell. It constitutes of top and bottom layers as flourine-doped tin oxide (FTO) and metal electrodes, respectively and electron transport layer (ETL) and hole transport layer (HTL) of TiO$_2$ (thickness = 50 nm) and Cu$_2$O (thickness = 50 nm), respectively. The choice of TiO$_2$ as ETL layer is owing to its appropriate energy level which is suitable for electron injection and hole blocking. TiO$_2$ thin film's conductance and mobility can further be enhanced by doping with various elements which tend to improve the device performance \cite{Guo2019}. Similarly, Cu$_2$O has been chosen as HTL because of its appropriate energy level alignment, high hole mobility, and longer lifetime of photo-excited carriers \cite{Yu2016}. Both the ETL and HTL layers play important role in PCE of perovskite solar cells. ETL and HTL provide low resistance path for charge carriers generated in the absorber layer. Consequently, doping concentration of ETL and HTL layer does not affects the performance of solar cell significantly. Further, with the increase of the thickness of ETL and HTL layers, the performance of solar cell also nearly remains constant \cite{Baig2018}. As few tens of nm of ETL and HTL give best performance results, we have chosen them to be 50 nm in thickness \cite{Patel2021, Minemoto2019}. Important material parameters attributed to the various ETL and HTL layers are similar to as reported by Patel et al. \cite{Patel2021} as shown in Table \ref{tbl:table1}. The active layer is the perovksite layer with composition CH$_3$NH$_3$PbI$_3$ and band gap of 1.55 eV while the thickness is varied. Here,the carrier mobility parameters have been taken from Hima et al. \cite{Hima2019}. We have chosen this composition and the bandgap as the perovskites around this bandgap show high spectral intensity and high stability under illumination \cite{Islam2021, Bush2017}. Further, the absorption coefficient data for the perovskite layer has been taken from the SCAPS-1D library. 


\begin{table}
  \includegraphics[width=\linewidth]{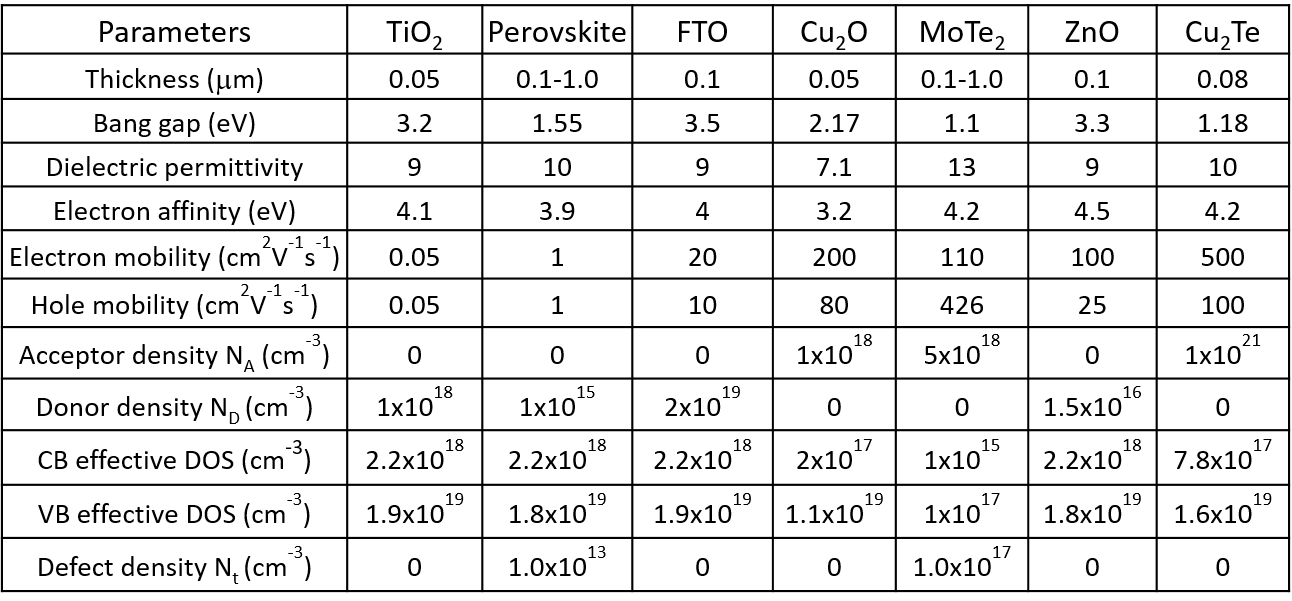}
  \caption{A table containing parameters of the various materials used in the simulation of single junction and 2T tandem solar cells.}
  \label{tbl:table1}
\end{table}

Similarly, Fig. \ref{fig:fig1}(b) shows arrangement of various layers in single junction solar cell containing MoTe$_2$. It constitutes of top and bottom layers as metal electrodes and ZnO (bandgap = 3.3 eV; thickness = 100 nm as taken from Hao et al. \cite{Hao2021}) and Cu$_2$Te (bandgap = 1.18 eV; thickness = 80 nm) as ETL and HTL/Back Surface Field layers, respectively. The active layer consists of MoTe$_2$ with the band gap 1.1 eV with the material parameters taken from Faisal et al. \cite{Faisal2017}. Further, the absorption coefficient data for the MoTe$_2$ layer has been taken to have square root dependence with respect to energy. Fig. \ref{fig:fig1}(c) shows arrangement of various layers in the pervoskite and MoTe$_2$ based 2T tandem solar cell. It constitutes of top and bottom subcells with as metal electrodes, ETL and HTL layers corresponding to the respective perovskite and MoTe$_2$ single junction solar cells. In addition, an interfacial layer such as any of TiO$_2$, ITO (Indium Tin oxide), and IZO (Indium Zinc oxide)  can be used between the two subcells \cite{Li2020}. This layer allows current of the opposite polarity charge carriers from the two subcells to recombine here. Therefore, a current density matching between the top and bottom subcell is required for 2T tandem configuration solar cells. 


\begin{figure}[h!]
 \includegraphics[width=1.0\textwidth]{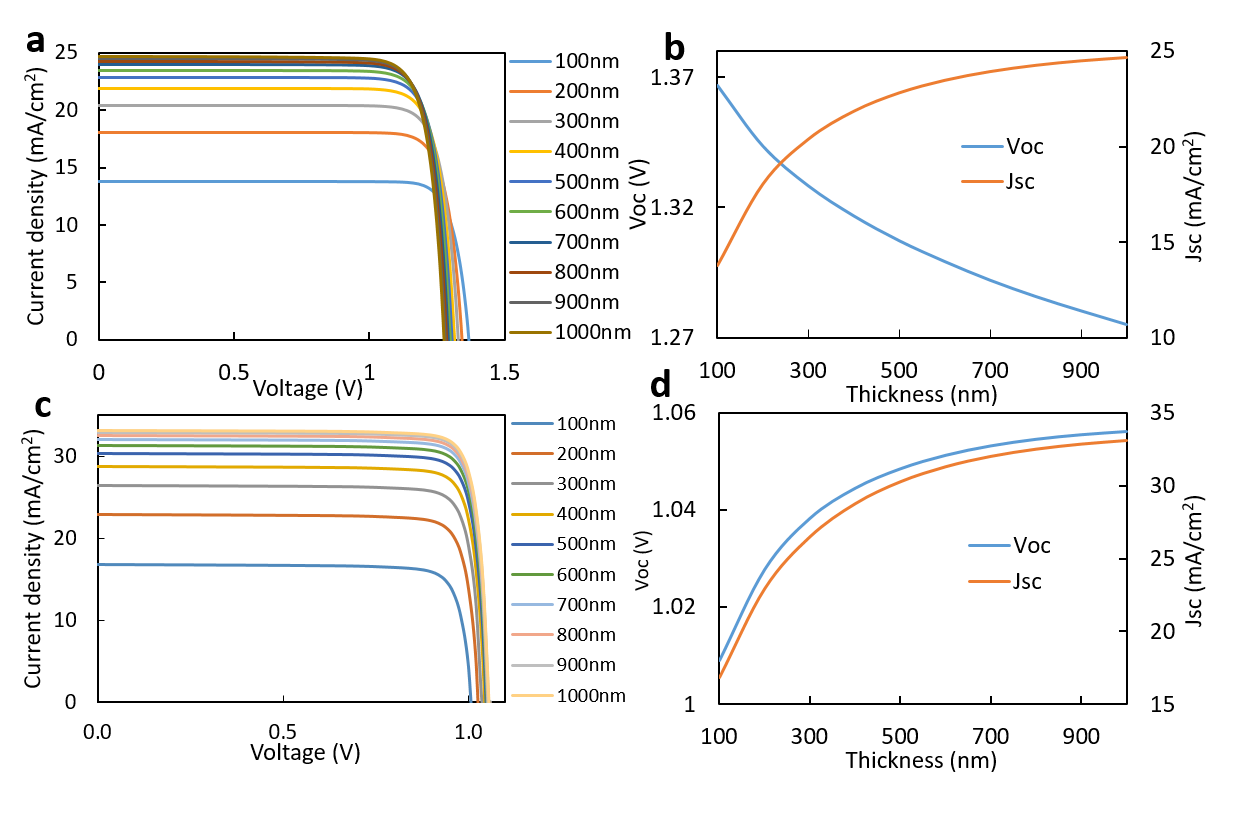}
 \caption{\textbf{a} shows calculated J-V curves for single junction CH$_3$NH$_3$PbI$_3$ solar cell under AM 1.5 G illumination for various thickness of the absorber layer. \textbf{b} shows variation of J$_{sc}$ and V$_{oc}$ as function of thickness of the perovskite layer. \textbf{c} shows calculated J-V curves under single junction MoTe$_2$ solar cell under AM 1.5 G illumination for various thickness of the MoTe$_2$ absorber layer. \textbf{d} shows variation of J$_{sc}$ and V$_{oc}$ values as function of thickness of MoTe$_2$ layer.}
  \label{fig:fig2}
\end{figure}

We carry out the simulation of the single junction solar cell containing CH$_3$NH$_3$PbI$_3$ layer using SCAPS-1D software under unfiltered AM 1.5 G illumination. SCAPS-1D is used to simulate and study various kind of single junction and tandem solar cells \cite{Patel2021, Adewoyin2019, Islam2021}. All the simulations have been performed at temperature 300 K. Further, the losses due to the reflection at interfaces as well as series resistance of the device have not been taken into consideration. Fig. \ref{fig:fig2}(a) show current density \textit{vs} voltage (J-V) curves for the perovskite single cell for various thickness of the absorber layer. The thickness of the absorber layer has been varied from 100 nm to 1000 nm while keeping other optimized parameters the same as in table \ref{tbl:table1}. We observe that with increasing thickness, the short circuit current density, J$_{sc}$, increases with the thickness of the perovskite layer while the value of open circuit voltage, V$_{oc}$, decreases as shown in Fig. \ref{fig:fig2}(b). As the thickness increases from 100 nm to 1000 nm, J$_{sc}$ gets nearly doubled while saturating at higher thickness and V$_{oc}$ decreases by 10 \% monotonically. The decrease in V$_{oc}$ may be attributed to higher recombination rates of the charge carriers with increasing thickness \cite{asd}. On the other hand, the initial increase in J$_{sc}$ is attributed to the increase in the active absorber volume which results in enhanced generation of charge carriers. But, with further increase of thickness, J$_{sc}$ stays the same due to enhanced layer resistance and carrier recombination. In this calculation, we have kept the acceptor doping concentration of the absorber layer to be 10$^{15}$ cm$^{-3}$. Higher doping density will decrease both the J$_{sc}$ as well as power conversion efficiency (PCE) of the solar cell due to enhanced recombination rate leading to decrease in the minority charge carriers \cite{Patel2021, Singh2021}. Further, we have assumed the interfaces between the perovskite and neighbouring layers to be defect free. 


Next, we perform simulation of the single junction solar cell containing MoTe$_2$ as absorber layer as shown in Fig. \ref{fig:fig1}(b) under unfiltered AM 1.5 G illumination. As earlier, the other parameters of the cell has been kept fixed as shown in Table \ref{tbl:table1} while the thickness of MoTe$_2$ is varied. Fig. \ref{fig:fig2}(c) shows J Vs V curves for various thickness of the absorber layer. Similar to the single perovskite solar cell, thickness of MoTe$_2$ layer seems to influence performance of solar cells significantly. Here too, the value of J$_{sc}$  nearly doubles as the thickness of the MoTe$_2$ layer is increased from 100 nm to 1000 nm. On the other hand, V$_{oc}$ is in the range of 1.02 V which is lower than the single junction perovskite solar cell due to the lower band gap of MoTe$_2$ layer. Even though the change in the V$_{oc}$ is significantly small, it increases with increasing thickness due to higher doping concentration of MoTe$_2$ layer. The increase in J$_{sc}$ initially and saturating at larger thickness is again due to enhanced increase in active absorber volume. Since, the saturation of J$_{sc}$ is related to charge carrier diffusion length, it tends to saturate at nearly 600 nm. The recombination rate will be significantly higher for the absorber layer having thickness more than 600 nm. 


\begin{figure}[h!]
 \includegraphics[width=1.0\textwidth]{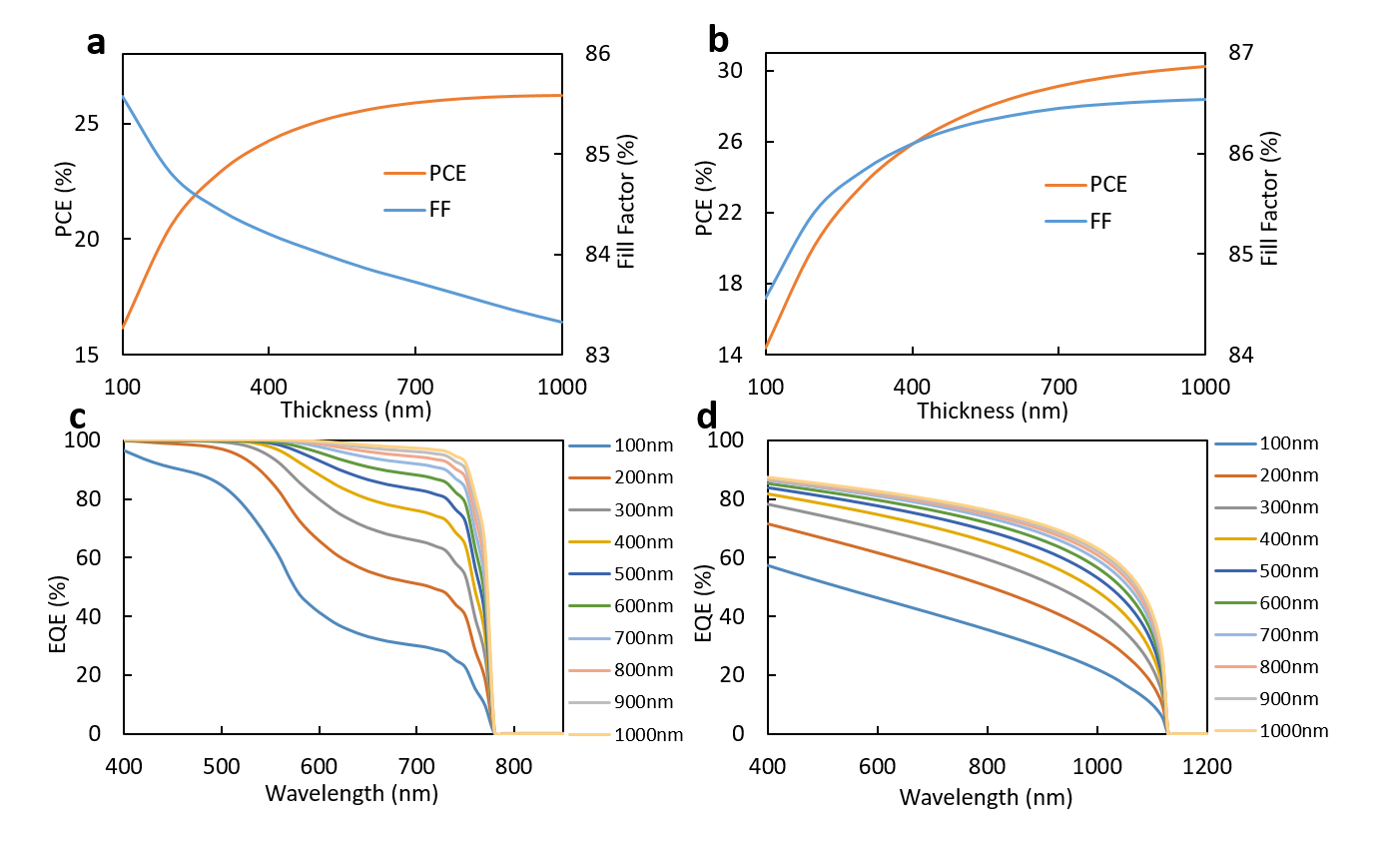}
  \caption{\textbf{a} shows dependence of PCE and FF on the thickness of CH$_3$NH$_3$PbI$_3$ absorber layer of the single junction solar cell. \textbf{b} shows dependence of PCE and FF on the thickness of MoTe$_2$ absorber layer of the single junction MoTe$_2$ solar cell. \textbf{c} shows dependence of EQE on the thickness of the perovskite absorber layer of the single junction perovskite solar cell. \textbf{d} shows dependence of EQE on the thickness of the MoTe$_2$ absorber layer of the single junction MoTe$_2$ solar cell.}
  \label{fig:fig3}
\end{figure}

As evident from Fig. \ref{fig:fig2}(a and c), the absorber layer thickness has strong effect on the Fill Factor (FF) and PCE of the two single junction solar cells as shown in Fig. \ref{fig:fig3}. We observe that the PCE of the perovskite solar increases as the thickness of the absorber layer increases and saturates to 25 \% at the thickness of $\sim$600 nm. This may be attributed to the enhanced absorption of photons leading to generation of charge carriers as the thickness of the abosrber layer increases. However for the thickness approximately more than 600 nm, the unwanted recombination starts to take over which leads to either saturation or the lowering of the PCE. The critical thickness of the absorber layer is related to the diffusion length of the charge carriers. This is consistent with the experimental demonstration of the PCE of the perovskite solar cell \cite{Wu2021}. The value of the FF decreases with the increasing thickness however the relative change is not significant.

The PCE of single junction MoTe$_2$ solar cell seems to increase with increasing thickness and saturates and 30 \% at 1000 nm as shown in Fig. \ref{fig:fig3}(b). We should note that there is no experimental demonstration of solar cell containing MoTe$_2$ as an absorber layer. Therefore, it will be interesting to experimentally realize solar cell based on MoTe$_2$ to measure its PCE and the other performance parameters. On the other hand, Memaran et al. \cite{Memaran2015} have reported experimental demonstration of single junction solar cells containing multilayer MoS$_2$ in field effect transistor (FET) geometry with efficiency 2.8 \%. Further experimental investigation is required to synthesize TMDC based solar cells and enhance their performance parameters. 


Figure \ref{fig:fig3}(c) shows dependence of external quantum efficiency (EQE) on the thickness of the absorber layer of the single junction perovskite solar cell. The shape of the EQE is attributed to variation of the refractive index of the perovskite layer on the wavelength of the light which affects the reflecting properties of the layer \cite{Brinkmann2021}. Further, as the thickness of the absorber layer is increased, EQE increases and acquires a step shape at higher thickness. The edge of the step at 800 nm marks 1.55 eV, the band gap of the CH$_3$NH$_3$PbI$_3$ layer used in the solar cell. Compared to single junction perovskite solar cell, EQE of the MoTe$_2$ solar cell (Fig. \ref{fig:fig3}(d)) shows slightly different dependence with the wavelength which may be attributed to the wavelength dependent refractive index of the material. However, EQE of MoTe$_2$ solar cell remains lower than that of the perovskite solar cell. Further, EQE of the MoTe$_2$ solar cell also increases with the increase in the thickness of the MoTe$_2$ layer with the step at 1125 nm which marks the band gap of MoTe$_2$. \cite{Ruppert2014}. Relatively lower EQE of MoTe$_2$ cell limits the current density of the bottom subcell in tandem geometry as discussed later.

\begin{figure}[h!]
 \includegraphics[width=1.0\textwidth]{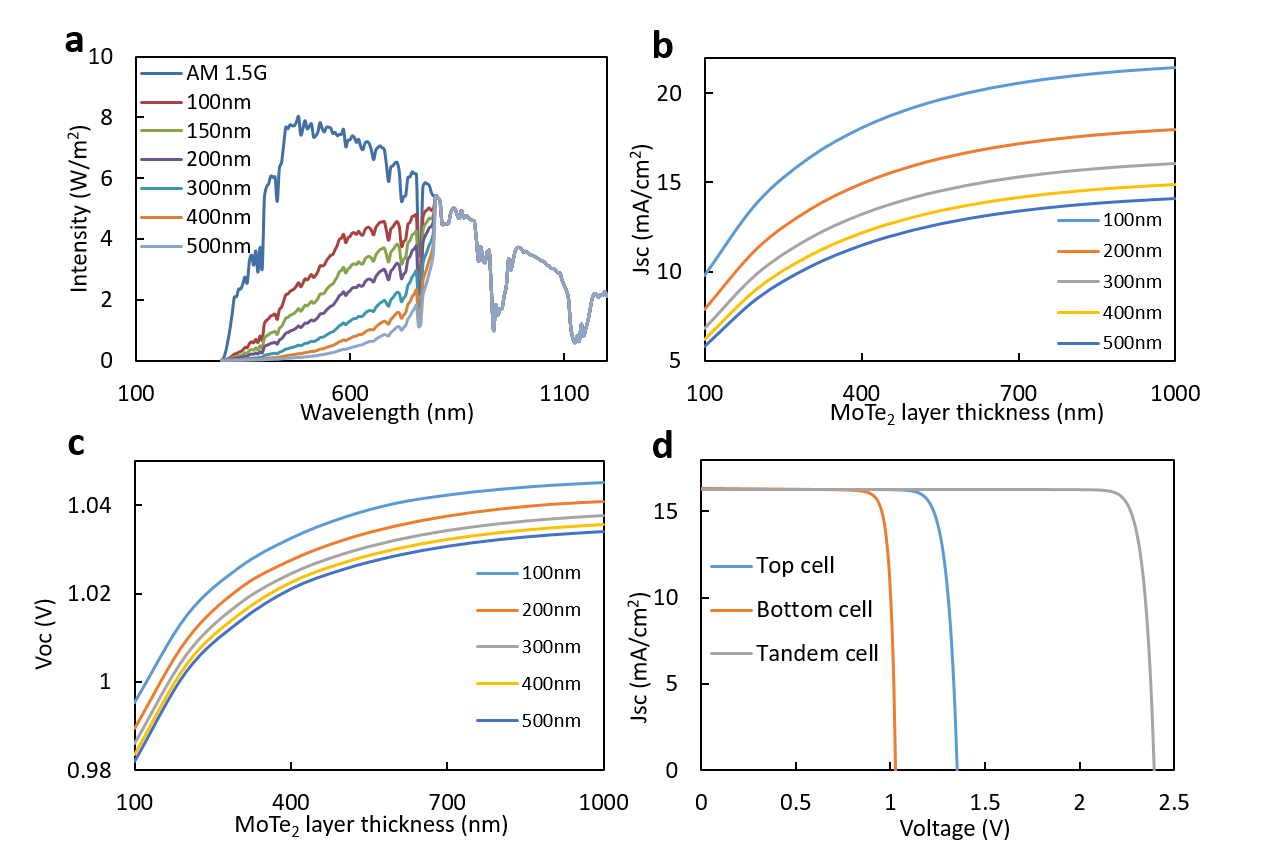}
  \caption{\textbf{a} shows the intensity of filtered AM 1.5G spectrum transmitted through the top subcell for various thickness of the perovskite layer of the top subcell. \textbf{b} shows variation of J$_{sc}$ as function of thickness of MoTe$_2$ bottom subcell for various thickness of the perovskite layer in the top subcell. Different thickness of perovskite layer results into different filtered power spectrum illuminating the bottom subcell. \textbf{c} shows variation of V$_{oc}$ as function of thickness of the MoTe$_2$ absorber layer for various thickness of the perovskite layer in the top subcell. \textbf{d} shows J-V plot for the current density matched tandem configuration for CH$_3$NH$_3$PbI$_3$ top and MoTe$_2$ bottom subcells and the tandem solar cell.} 
  \label{fig:fig5}
\end{figure}

As SCAPS-1D software (version 3.3.08) does not support simulation of 2T tandem solar cell structures, we employ the general technique where the top subcell is analyzed first using standard, unfiltered AM 1.5G spectrum and the filtered spectrum after the top subcell is used to illuminate the bottom subcell \cite{Madan2020,Islam2021}. Figure \ref{fig:fig5}(a) shows power dependence of the transmitted spectrum on the thickness of CH$_3$NH$_3$PbI$_3$ layer following the top subcell. Above the wavelength higher than 800 nm corresponding to the band gap of CH$_3$NH$_3$PbI$_3$ layer, the active layer remains transparent and all the power gets transmitted. Further, between 500 nm and 800 nm wavelength, we see that the transmitted power varies with the thickness of the absorber layer depending on amount of active volume available for absorption. We use this filtered spectrum to illuminate the bottom subcell and perform further calculations. Fig. \ref{fig:fig5}(b and c) shows the variation of J$_{sc}$ and V$_{oc}$ extracted from J-V curves as function of thickness of the MoTe$_2$ layer as illuminated by the filtered spectrum for various thickness of top subcell absorber layer. As expected, for a higher thickness of the top subcell active layer, the filtered spectrum will have lower intensities and therefore it will give rise to lower values of J$_{sc}$ and V$_{oc}$ for the bottom subcell. This indicates a trade-off between the current densities of the two subcells against their thickness. Further, for the tandem solar cell, the final current density is limited by the smaller of the two, therefore, we choose appropriate thickness of the two cells to achieve a current density matching. Fig. \ref{fig:fig5}(d) shows J-V plot for the current density matched top subcell, bottom subcell and 2T tandem solar cell. The current matching condition has been achieved for thickness of 150 nm and 407 nm for CH$_3$NH$_3$PbI$_3$ and MoTe$_2$ layers, respectively which correspond to the current density of 16.3 mA/cm$^2$. Here, the final V$_{oc}$ is 2.4 V and the FF is 90.3. The calculated efficiency is 35.3 \% which is significantly higher than the individual single junction solar cells. For comparative analysis, we have also investigated 2T tandem solar cells containing WS$_2$ and WSe$_2$ as bottom subcells (see Supplementary Information Table S1). Among all the tandem solar cell proposed, tandem structure with perovskite and MoTe$_2$ as the absorber layers of top and bottom subcells show the highest efficiency of 35.3 \%.

\begin{table}
\caption{Various performance parameters such as J$_{sc}$, V$_{oc}$, FF, and PCE for various single junction pervoskite (PVK) and TMDC solar cells for optimized material parameters and absorber layer thickness of 1000 nm. Similarly, various performance parameters for tandem solar cell (PVK/TMDC) consisting of perovskite and TMDCs as top and bottom subcells, respectively for current density matched  thicknesses. }
\centering
\begin{tabular}{ | m{2.2cm} | m{2.5cm}| m{2.65cm} | m{1.4cm}| m{2.5cm} | m{1.3cm}| m{1.0cm} |} 
\hline
  & bandgap (eV) & thickness (nm) & V$_{oc}$ (V) & J$_{sc}$ (mA/cm$^2$) & FF (\%) & PCE (\%) \\ 
\hline
PVK & 1.55 & 1000 & 1.27 & 24.7 & 83.3 & 26.2\\ 
\hline
MoTe$_2$ & 1.1 & 1000 & 1.05 & 33.1 & 86.5 & 30.3\\ 
\hline
PVK/MoTe$_2$ & 1.55/1.1 & 150/407 & 2.4 & 16.3 & 90.3 & 35.3\\ 
\hline
\end{tabular} 
\label{Tab:table2}
\end{table}

\section*{Conclusions}

We have proposed various 2T tandem solar cells consisting of CH$_3$NH$_3$PbI$_3$ and TMDCs as absorber layers of the top and bottom subcells and have analysed their performance parameters by using SCAPS-1D software package. Single junction perovskite and MoTe$_2$ solar cells have maximum PCE of 26.2 \% and 30.3 \%, respectively for the optimized material parameters and layer thicknesses of 1000 nm. On the other hand 2T tandem structure can achieve PCE of 35.3 \% which is significantly higher than either single junction solar efficiencies. Such high PCEs indicate that perovskite and TMDCs based solar cells are interesting for experimental investigation and carry the potential to be used for manufacturing on flexible substrates through roll-to-roll based technologies.

\begin{suppinfo}
Two-Terminal Tandem Solar Cells based on Perovskite and Transition Metal Dichalcogenides
\end{suppinfo}

\begin{acknowledgement}

We thank Dr. Rajan Jha, IIT Bhubaneswar for discussions. This research was supported by the Department of Science and Technology (SERB -- CRG/2020/004293). 
Further, the authors are thankful to Marc Burgelman and his team at the Electronics and Information Systems (ELIS), University of Gent, Belgium for providing access to SCAPS-1D software package. The authors declare no competing financial interest. 

\end{acknowledgement}

\bibliography{manuscript}

\end{document}